\documentstyle[aps,preprint,epsf]{revtex}
\begin{document}

\tighten

\title
{\bf Optical Conductivity in a Simple Model of Pseudogap State in
Two-Dimensional System}
\author{M.V.Sadovskii}
\address
{Institute for Electrophysics,\\ Russian Academy of Sciences,\ Ural Branch,\\
Ekaterinburg,\ 620049, Russia\\
E-mail:\ sadovski@ief.uran.ru} 
\maketitle

\begin{abstract}
We present calculation of optical conductivity in a simple model of electronic
spectrum of two-dimensional system with ``hot patches'' on the Fermi surface,
leading to non Fermi-liquid renormalization of the spectral density 
(pseudogap) on these patches. It is shown that this model qualitatively
reproduces basic anomalies of optical experiments in the pseudogap state of
copper oxides.
\end{abstract}

\newpage

Among the number of anomalies of the normal phase of high-temperature
superconductors especially interesting is the observation of the pseudogap
in the electronic spectrum of underdoped copper oxides\cite{Ran,RC}.  
Most striking evidence for this unusual state was obtained in the
ARPES experiments\cite{Ding,D}, which demonstrated the anisotropic evolution
of electron spectral density. In particular, in these experiments the
maximal value of the pseudogap was observed in the vicinity of ($\pi,0$) 
point in the Brillouin zone, while in the direction of diagonals of the zone
the pseudogap was absent. Accordingly, around ($\pi,0$) the Fermi surface is
completely destroyed, while around diagonals it is conserved. In this sense
it is usually said that the symmetry of the pseudogap is of ``$d$-wave'' type
and coincides with the symmetry of the superconducting gap in these compounds.
These anomalies are observed up to the temperatures $T\simeq T^*$, which are
significantly higher than the temperature of superconducting transition $T_c$.

Pseudogap effects are also observed as certain anomalies of the optical
conductivity of the number of high-temperature superconductors
\cite{B1,B2,B3,B4,B5}. These anomalies are manifested mainly in the
existence of anomalously narrow Drude--like peak (the drop in effective
scattering rate) in the region of small frequencies and rather weak
absorption through pseudogap at higher frequencies.

There is a number of theoretical approaches, trying to explain pseudogap
anomalies. In this paper we assume that these anomalies are mainly due to
fluctuations of antiferromagnetic short-range order, as in the ``hot-spots''
model\cite{Sch,SchP}. In this model it is possible to obtain ``nearly'' exact
solution for the electronic spectrum, based upon complete summation of all
the relevant Feynman diagrams, describing electron interaction with
antiferromagnetic fluctuations\cite{Sch,SchP,KS}, generalizing to two
dimensions the earlier solution of a similar one-dimensional problem
\cite{C1,C2,C79,C91,C91a}.

Calculations in ``hot spots'' model are complicated by the use of
``realistic'' spectrum of current carriers, so that in this work we shall
consider much simplified ``hot patches'' model of the pseudogap state 
proposed in Ref.\cite{PS}, which is physically quite close to ``hot spots''
model. Following Ref.\cite{PS} we assume that the Fermi surface of 
two-dimensional electronic system is like shown in Fig.1. Analogous model
of the Fermi surface was considered in Ref.\cite{Z}, where it was stressed,
that it is quite close to that observed in a number of high-temperature
superconductors.  Fluctuations of short-range order are assumed to be static
and Gaussian with correlation function of the form\cite{KS,PS}:  
\begin{equation} 
S({\bf q})=\frac{1}{\pi^2} 
\frac{\xi^{-1}}{(q_x-Q_x)^2+\xi^{-2}}
\frac{\xi^{-1}}{(q_y-Q_y)^2+\xi^{-2}}
\label{fluct}
\end{equation}
where either $Q_x=\pm 2p_F$,\ $Q_y=0$ or $Q_y=\pm 2p_F$,\ $Q_x=0$.
We shall assume that these fluctuations interact only with electrons from
the ``hot'' (flat) patches of the Fermi surface shown in Fig.1.
Effective interaction of these electrons with fluctuations we shall model as
$(2\pi)^2\Delta^2S({\bf q})$, where $\Delta$ is of dimensions of energy and
defines the characteristic width of the pseudogap \footnote{More formally we
may say that we introduce an effective interaction ``constant'' of electrons
with fluctuations as:
$\Delta_{\bf p}=\Delta[\theta(p_x^0-p_x)\theta(p_x^0+p_x)+
\theta(p_y^0-p_x)\theta(p_y^0+p_x)]$.}.  
Thus, in fact we assume that scattering by fluctuations is of essentially
one-dimensional nature. The choice of scattering vector 
${\bf Q}=(\pm 2p_F,0)$ or ${\bf Q}=(0,\pm 2p_F)$ corresponds to the picture
of incommensurate fluctuations. Commensurate case with
${\bf Q}=(\frac{\pi}{a},\frac{\pi}{a})$ (where $a$ - is lattice constant) 
can also be analyzed. 

On ``cold'' patches we shall assume the existence of some weak static
scattering of rather arbitrary nature with appropriate scattering rate
described by phenomenological parameter $\gamma$, with $\gamma\ll \Delta$
in most cases, so that on ``hot'' patches it can be just neglected.
Accordingly on these ``cold'' patches the electronic spectrum is described
by the usual Green's function for the system with weak scattering
(Fermi--liquid).

In the limit of $\xi\rightarrow \infty$ this model can be solved exactly
by method used in Refs.\cite{C1,C2}, while for finite $\xi$ it can be
``nearly'' exactly solved (cf.\cite{Sch,SchP,KS}) by the method of Refs.
\cite{C79,C91,C91a}.

At first we shall consider maximally simplified case of $\xi\rightarrow\infty$, 
when an effective interaction with fluctuations (\ref{fluct}) takes the
simplest form:
\begin{equation} 
(2\pi)^2\Delta^2\left\{\delta(q_x\pm 2p_F)\delta(q_y)+
\delta(q_y\pm 2p_F)\delta(q_x)\right\}
\label{WW}
\end{equation}
In this case we can easily sum all perturbation series for an electron
scattered by these fluctuations using the method of Refs.\cite{C1,C2}, 
both for one-electron and two-electron Green's functions. For the case of
incommensurate fluctuations of short-range order the one-electron Green's
function becomes:
\begin{equation} 
G(\epsilon,{\bf p})=\int \limits_0^\infty 
d{\zeta}e^{-\zeta} 
\frac{\epsilon+\xi_{\bf p}}{\epsilon^2-\xi_{\bf p}^2-\zeta \Delta^2(\phi)},
\label{fgrina}
\end{equation}
where $\xi_{\bf p}=v_F(|{\bf p}|-p_F)$ ($v_F$--Fermi velocity) and 
$\Delta(\phi)$ is defined for $0\leq\phi\leq\frac{\pi}{2}$ as:  
\begin{equation}
\Delta(\phi)=\left\{
\begin{array}{ll}
\Delta & ,0\leq\phi\leq\alpha,\>\frac{\pi}{2}-\alpha\leq\phi\leq\frac{\pi}{2} 
\\ 0 & ,\alpha\leq\phi\leq\frac{\pi}{2}-\alpha 
\end{array} \right.  
\label{D} 
\end{equation}
where $\alpha=arctg(\frac{p_y^0}{p_F})$, $\phi$ - is polar angle, defining 
the direction of the vector ${\bf p}$ in ($p_x,p_y$) plane. For the other
values of $\phi$ parameter $\Delta(\phi)$ is defined by the obvious symmetry
considerations similarly to (\ref{D}). It is easily seen that changing 
$\alpha$ in the interval of $0\leq\alpha\leq\frac{\pi}{4}$, we in fact
change the size of ``hot'' patches on the Fermi surface, where ``nesting''
condition $\xi_{\bf{p-Q}}=-\xi_{\bf p}$ is satisfied. In particular, 
$\alpha=\pi/4$ corresponds to a square Fermi surface, where the ``nesting''
condition is satisfied everywhere. Below we always rather arbitrarily assume 
$\alpha=\pi/6$. Qualitative dependencies of a number of physical properties
of the model on $\alpha$ are given in Ref.\cite{PS}. Outside ``hot'' patches
(second inequality in (\ref{D})) Green's function (\ref{fgrina}) just
coincides with a free one (in fact here we must only take into account the
above mentioned weak scattering $\gamma$).

Spectral density and density of states, defined by Green's function
(\ref{fgrina}) were given in Ref.\cite{PS} and demonstrate non Fermi-liquid
(pseudogap) behavior on ``hot'' patches of the Fermi surface. Two-particle
Green's function (density-density response function) on ``hot'' patches can
be calculated also by complete summation of appropriate diagrams as it was
done before in one-dimensional case\cite{C1,C2}.

Conductivity in this model is determined by additive contributions from
``hot'' and ``cold'' patches. For its real part we obtain:

\begin{equation}
Re\sigma(\omega)=\frac{4\alpha}{\pi}Re\sigma_{\Delta}(\omega)
+(1-\frac{4\alpha}{\pi})Re\sigma_D(\omega)
\label{Resigma}
\end{equation}
where with the account of results of Ref.\cite{C1,C2} we have:
\begin{equation}
Re\sigma_{\Delta}(\omega)=\frac{\omega_p^2}{4}\frac{\Delta}
{\omega^2}\int \limits_0^{\omega^2/4\Delta^2}
d{\zeta} exp(-\zeta)\frac{\zeta}{\sqrt{\omega^2/4\Delta^2-\zeta}}
\label{Dsig}
\end{equation}
where $\omega_p$--plasma frequency and
\begin{equation}
Re\sigma_D(\omega)=\frac{\omega_p^2}{4\pi}\frac{\gamma}{\omega^2+\gamma^2}
\label{Drude}
\end{equation}
--is the usual Drude--like conductivity from ``cold'' patches.

In Fig.2 we present the frequency dependence of the real part of conductivity,
calculated from (\ref{Resigma}), (\ref{Dsig}), (\ref{Drude}) for different
values of $\gamma$. Even in this simplest approximation these dependencies
are very similar to those observed in the experiments of Refs.
\cite{B1,B2,B3,B4,B5}. As the scattering rate $\gamma$ on ``cold'' patches
grows, the Drude--like peak at small frequencies is dumped.

More realistic case of finite correlation length of ``antiferromagnetic''
short-range order fluctuations $\xi$ in (\ref{fluct}) can be analyzed by the
method of Refs.\cite{C79,C91,C91a}, which allows to find ``nearly exact''
\cite{KS} solution of the problem. For one-electron Green's function on
``hot'' patches we obtain the following recurrence relation (continuous
fraction representation)\cite{C79}): 
\begin{equation}
G^{-1}(\epsilon,\xi_{\bf p})=G_{0}^{-1}(\epsilon,\xi_{\bf p})-
\Sigma_{1}(\epsilon,\xi_{\bf p})
\label{G}
\end{equation}
where
\begin{equation}
\Sigma_{k}(\epsilon,\xi_{\bf p})=\Delta^2\frac{v(k)}
{\epsilon-(-1)^k\xi_{\bf p}+ikv_F\kappa-\Sigma_{k+1}(\epsilon,\xi_{\bf p})}
\label{rec}
\end{equation}
Combinatorial factor (determining the number of diagrams):  
\begin{equation}
v(k)=\left\{\begin{array}{cc}
\frac{k+1}{2} & \mbox{for odd $k$} \\
\frac{k}{2} & \mbox{for even $k$}
\end{array} \right.
\label{vincomm}
\end{equation}
for the case of incommensurate fluctuations of short-range order.
For commensurate case:
\begin{equation} 
v(k)=k 
\label{vcomm}
\end{equation}
In spin-fermion model\cite{Sch,SchP}:
\begin{equation}
v(k)=\left\{\begin{array}{cc}
\frac{k+2}{3} & \mbox{for odd $k$} \\
\frac{k}{3} & \mbox{for even $k$}
\end{array} \right.
\label{vspin}
\end{equation}
For the vertex-part, determining density-density response function
(two-particle Green's function) on ``hot'' patches, we have the following
recurrence relation (details can be found in\cite{C91,C91a} and in
\cite{SchP}):  
\begin{eqnarray} 
J^{RA}_{k-1}(\epsilon,\xi_{\bf p};\epsilon+\omega,\xi_{\bf p+q})=\\ \nonumber
=e+\Delta^2v(k)G^A_k(\epsilon,\xi_{\bf p})G^R_k(\epsilon+\omega,\xi_{\bf 
p+q})J^{RA}_{k}(\epsilon,\xi_{\bf p};\epsilon+\omega,\xi_{\bf p+q})\times \\
\nonumber
\times\Biggl\{1+\frac{2iv_F \kappa k}{\omega-(-1)^k v_Fq+v(k+1)\Delta^2
[G^A_{k+1}(\epsilon,\xi_{\bf p})-G^R_{k+1}(\epsilon+\omega,\xi_{\bf p+q})]} 
\Biggr\}
\label{vertex}
\end{eqnarray}
where $e$--electronic charge, $R(A)$ denote retarded (advanced) Green's
function. Appropriate contribution of ``hot'' patches to conductivity
$Re\sigma_{\Delta}(\omega)$ in (\ref{Resigma}) can be calculated as in  
Ref.\cite{C91,C91a}, while $Re\sigma_D(\omega)$ is again given by(\ref{Drude}).
Typical results of these calculations are presented in Figs.3-6.
It is seen that difference between the results for incommensurate and
spin-fermion combinatorics are rather small. The general qualitative
picture is also conserved in the case commensurate combinatorics.
Real part of conductivity is characterized by rather narrow Drude-like peak
for small frequencies $\omega < \gamma$ due to ``cold'' patches on the Fermi
surface and relatively flat maximum for frequencies $\omega\sim 2\Delta$, 
corresponding to the absorption through the pseudogap which opens on ``hot''
patches. Drude-like peak is dumped with the growth of $\gamma$, while the
maximum at small frequencies, which can be seen in Fig.2 and Fig.3, can be
attributed to the ``remains'' of one-dimensional localization\cite{C91,C91a}.  
The dependence of conductivity on correlation length of fluctuations
$\xi=\kappa^{-1}$ is rather weak for all (most interesting) values of
parameters analyzed here. The qualitative picture obtained is very similar
to experimental data obtained for the number of high-temperature
superconducting copper oxides studied in Refs.\cite{B1,B2,B3,B4,B5}. 
Apparently there will be no problem with quantitative fitting of experimental
data using the known values of $\omega_p\sim 1.5-2.5eV$ and $2\Delta\sim 
0.1eV$, as well as the values of $\gamma$, which can be determined from the
width of the observed Drude-like peak, and varying ``free'' parameters
$\alpha$ (the size of ``hot'' patches) and $\xi$ (for this we also can use
the estimates from other experiments\cite{SchP}).

The author is grateful to Dr.E.Z.Kuchinskii for useful discussions. 

This work was supported in part by the grant of Russian Foundation of Basic
Research $N^o$ 99-02-16285 and also by the State Program ``Statistical
Physics'' as well as by the project $N^o$ 96-051 of the State Program on
HTSC of the Russian Ministry of Science.

\newpage

\begin{figure}
\epsfxsize=12cm
\epsfysize=21.5cm
\epsfbox{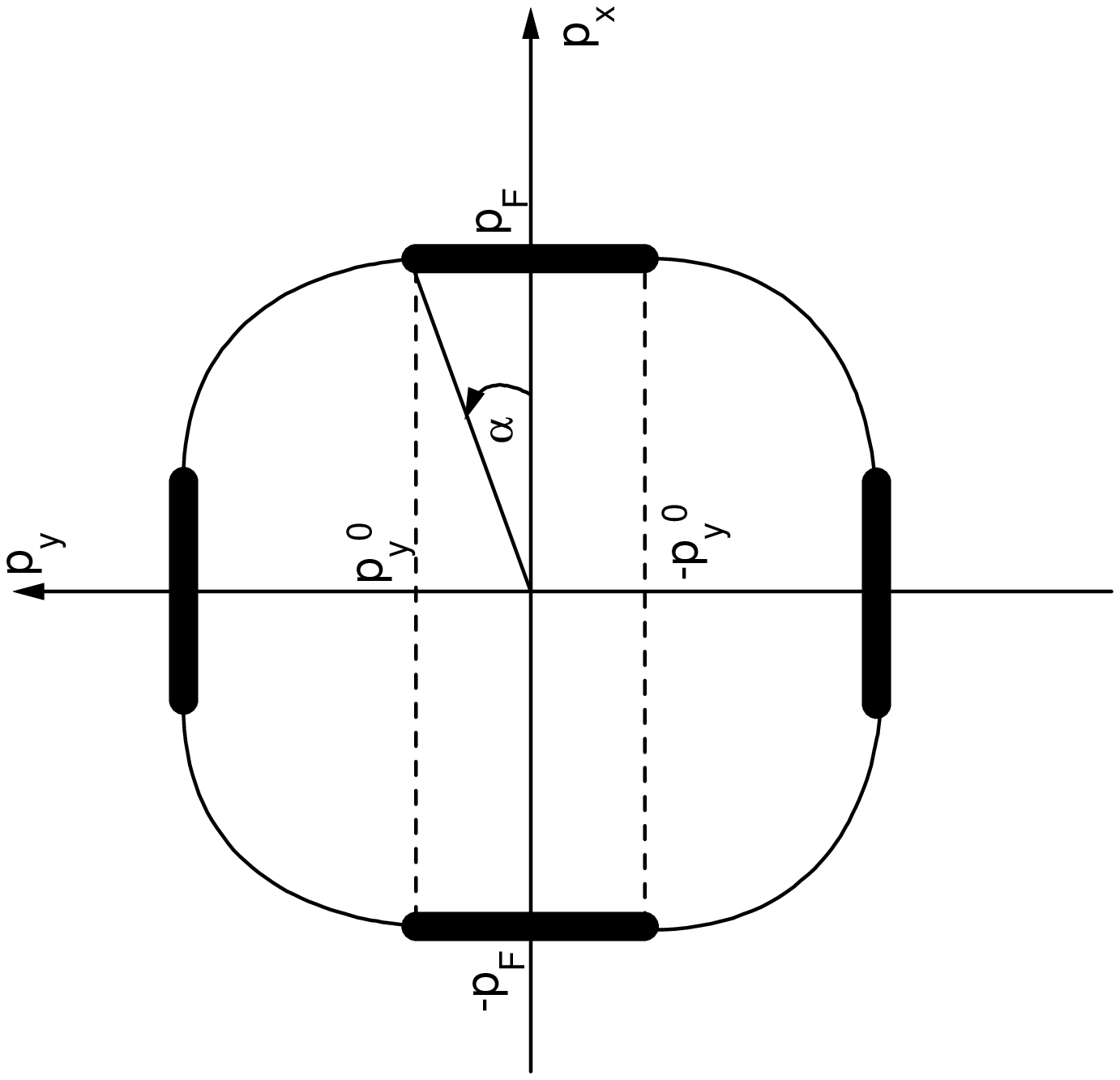}
\caption{Model Fermi--surface of two-dimensional system. ``Hot'' patches are
shown by thick lines with the width of the order of $\sim \xi^{-1}$. }
\end{figure}

\begin{figure}
\epsfxsize=16cm
\epsfysize=20cm
\epsfbox{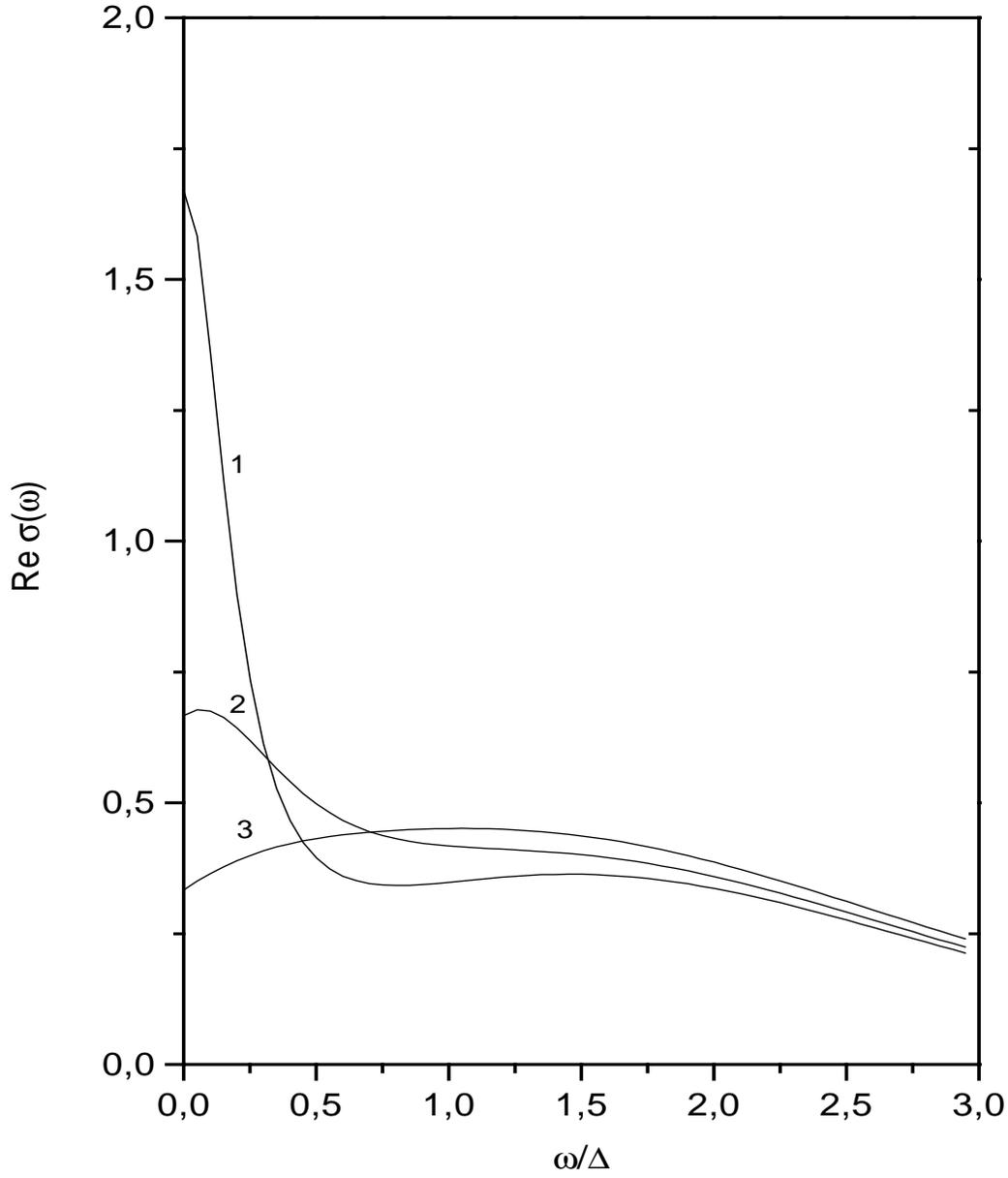}
\caption{Real part of conductivity in the model with infinite correlation length.
Conductivity is in units of $\omega_p^2/4\pi\Delta$. Incommensurate 
fluctuations.
(1)---$\gamma/\Delta=0.2$;\ (2)---$\gamma/\Delta=0.5$;\ 
(3)---$\gamma/\Delta=1.0$. }
\end{figure}

\begin{figure}
\epsfxsize=16cm
\epsfysize=20cm
\epsfbox{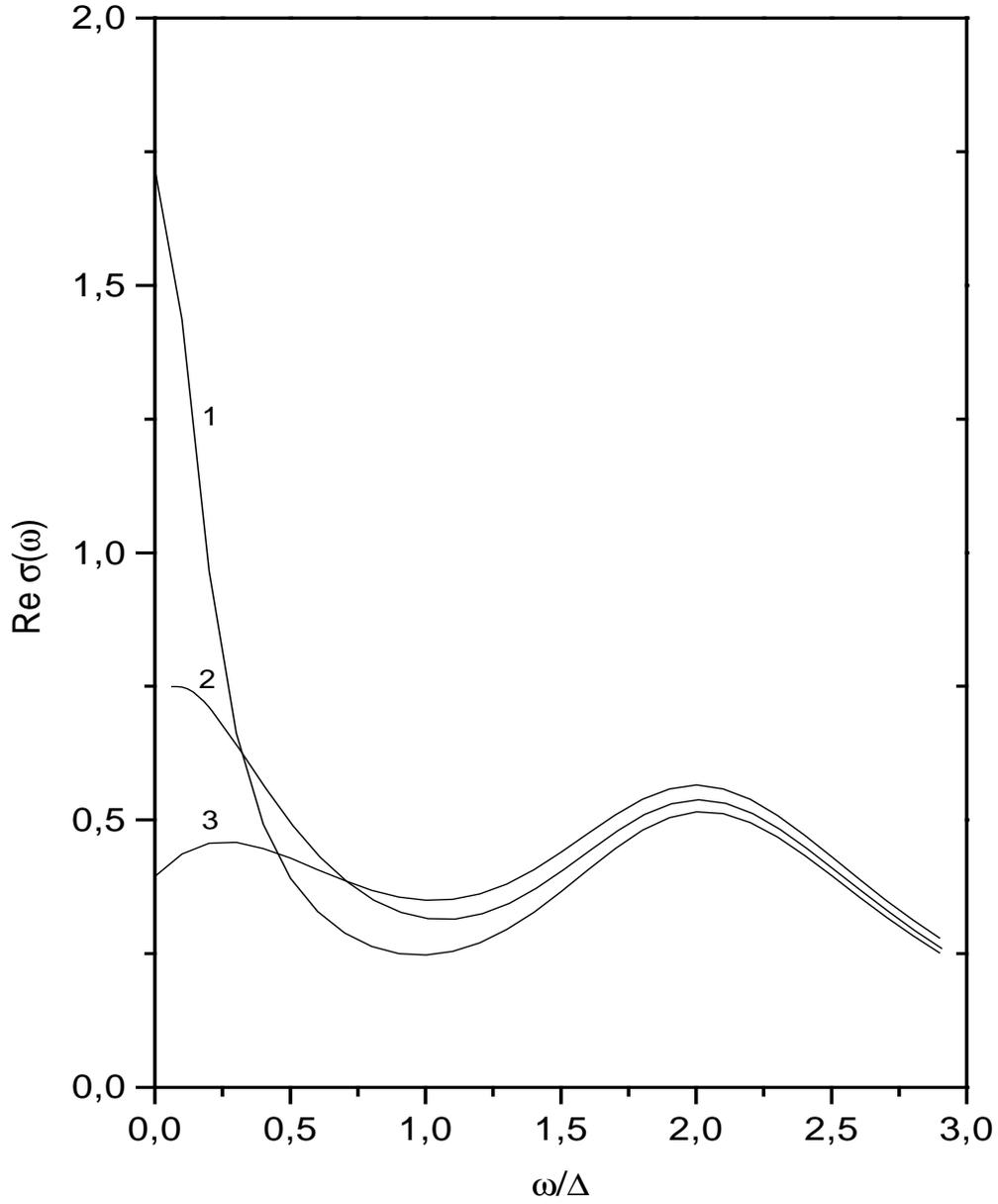}
\caption{Real part of conductivity as a function of $\gamma$ for the fixed
value of correlation length $v_F\kappa=0.5\Delta$.\ Incommensurate
fluctuations.
Conductivity is in units of $\omega_p^2/4\pi\Delta$.
(1)---$\gamma/\Delta=0.2$;\ (2)---$\gamma/\Delta=0.5$;\ 
(3)---$\gamma/\Delta=1.0$. } 
\end{figure}

\begin{figure}
\epsfxsize=16cm
\epsfysize=20cm
\epsfbox{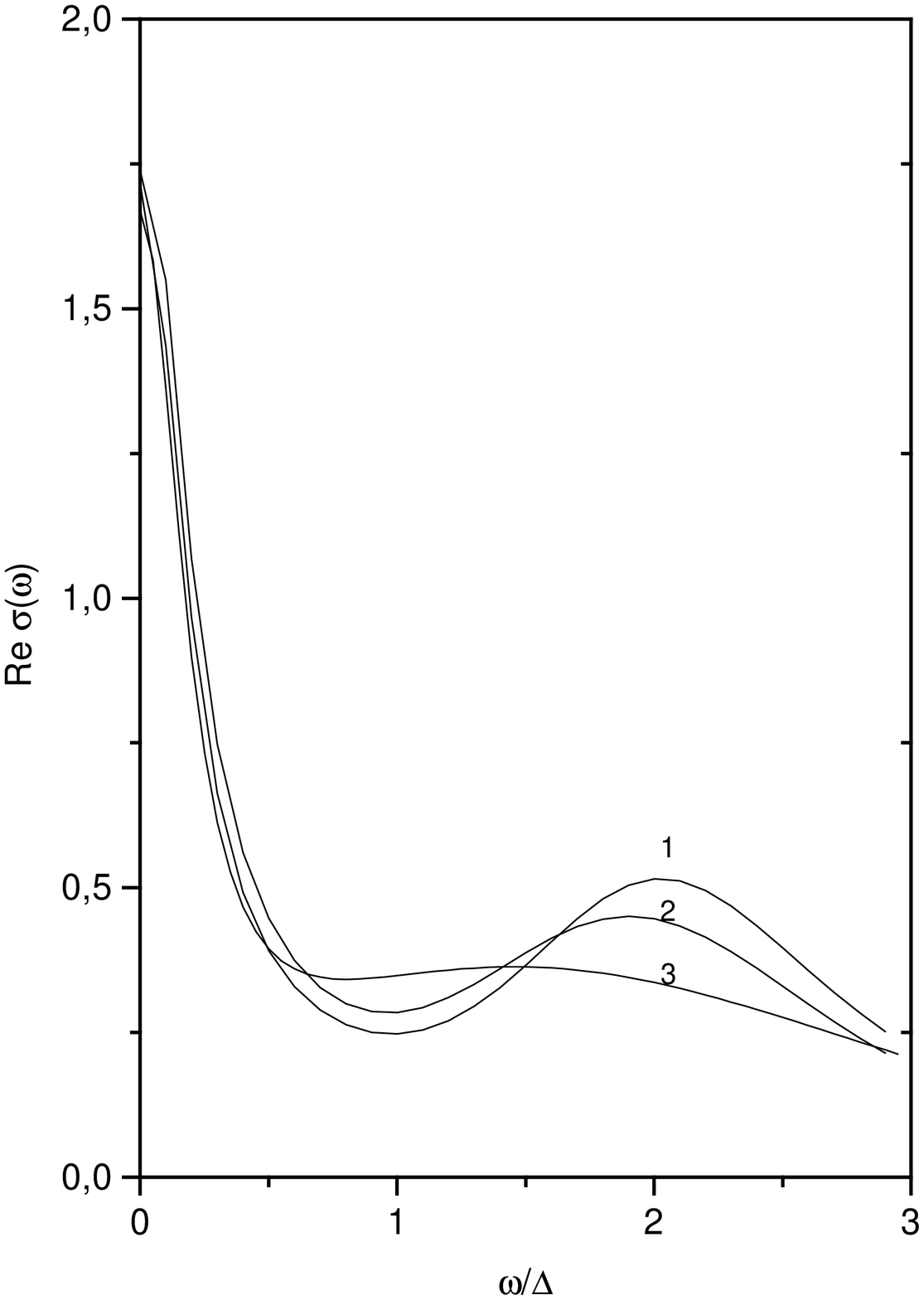}
\caption{Real part of conductivity as a function of correlation length for the
fixed value of $\gamma=0.2\Delta$.\ Incommensurate fluctuations.
Conductivity is in units of $\omega_p^2/4\pi\Delta$.
(1)---$v_F\kappa/\Delta=0.5$;\ (2)---$v_F\kappa/\Delta=1.0$;\ 
(3)---$v_F\kappa/\Delta=0$. }
\end{figure}

\begin{figure}
\epsfxsize=16cm
\epsfysize=20cm
\epsfbox{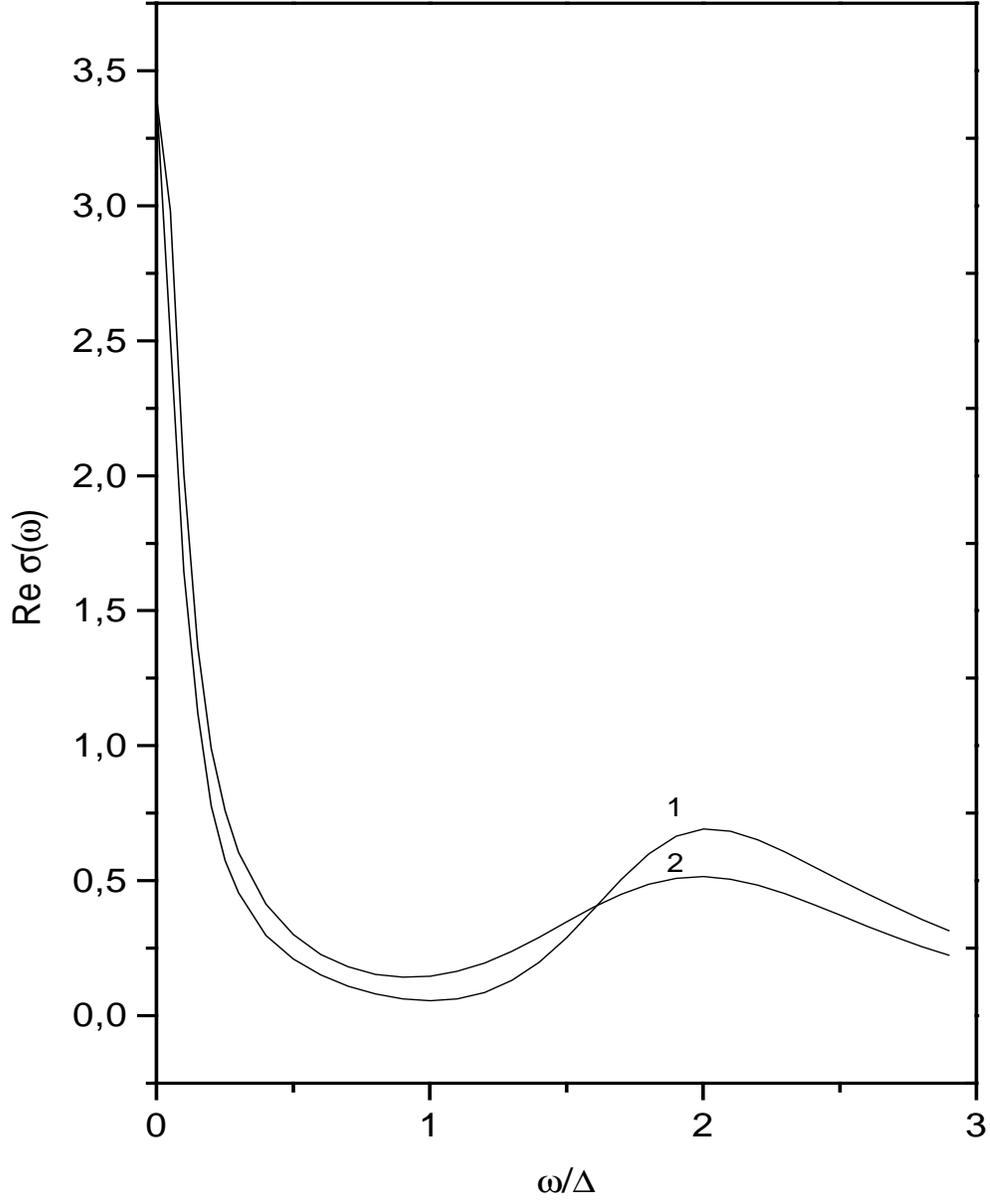}
\caption{Real part of conductivity as a function of correlation length for the
fixed value of $\gamma=0.1\Delta$.\ The case of spin-fluctuation model.
Conductivity is in units of $\omega_p^2/4\pi\Delta$.
(1)---$v_F\kappa/\Delta=0.5$;\ (2)---$v_F\kappa/\Delta=1.0$. }
\end{figure}

\begin{figure}
\epsfxsize=16cm
\epsfysize=20cm
\epsfbox{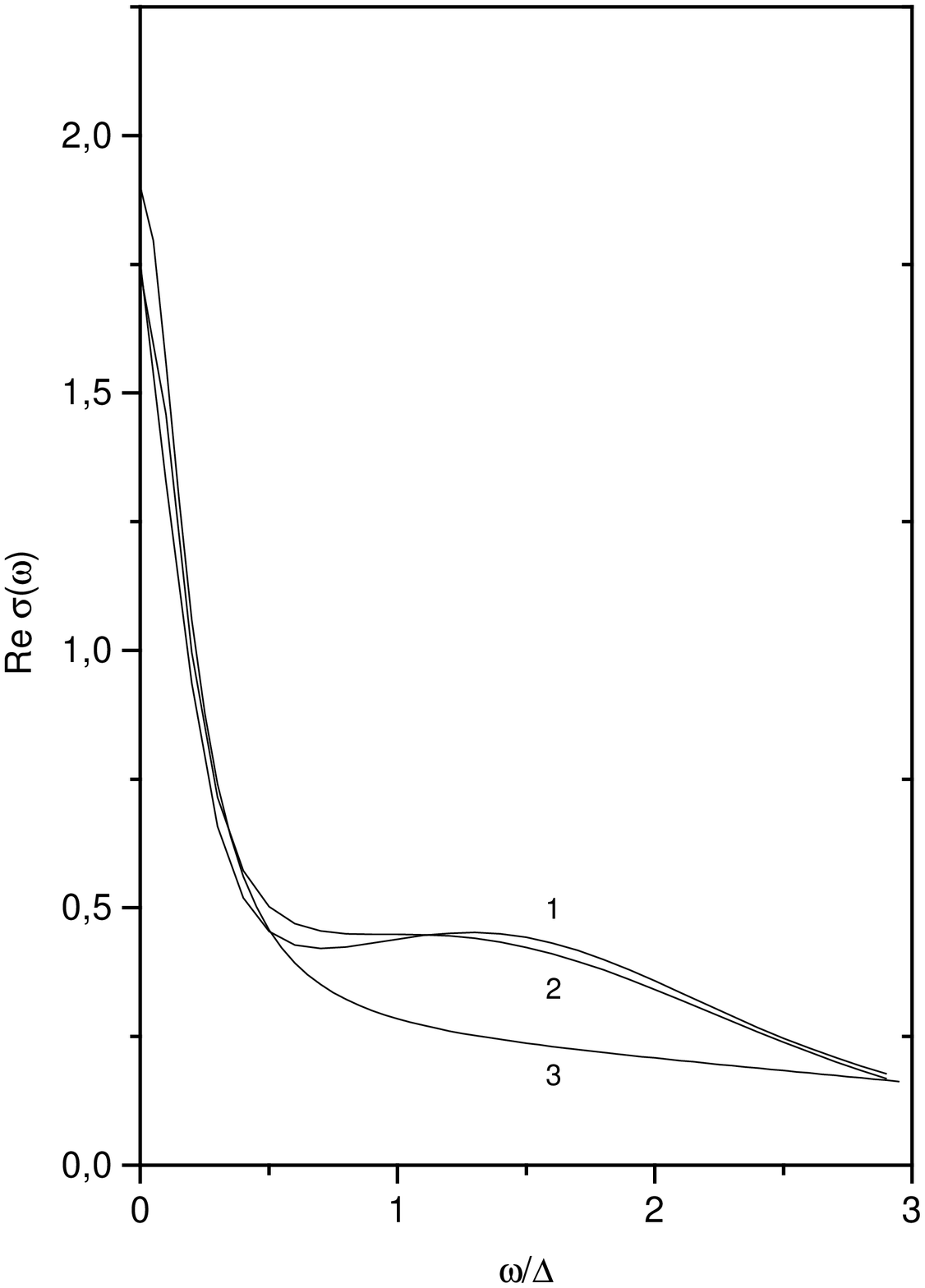}
\caption{Real part of conductivity as a function of correlation length for the
fixed value of $\gamma=0.2\Delta$.\ Commensurate fluctuations.
Conductivity is in units of $\omega_p^2/4\pi\Delta$.
(1)---$v_F\kappa/\Delta=0.5$;\ (2)---$v_F\kappa/\Delta=1.0$;\ 
(3)---$v_F\kappa/\Delta=0$. }
\end{figure}

\end{document}